\documentclass[final]{aipproc}

\layoutstyle{6x9}

\newcommand{\beq}[1]{\begin{equation}\label{#1}}
\newcommand{\eeq}{\end{equation}}
\newcommand{\bea}[1]{\begin{eqnarray}\label{#1}}
\newcommand{\eea}{\end{eqnarray}}

\def\ktr{\tilde{\kappa}_{\rm tr}}
\def\etal{{\it et al.}}

\def\lsim{\mathrel{\rlap{\lower3pt\hbox{$\sim$}}
    \raise2pt\hbox{$<$}}}
\def\gsim{\mathrel{\rlap{\lower3pt\hbox{$\sim$}}
    \raise2pt\hbox{$>$}}}

\def\EVCR{E_{\rm VCR}}
\def\GVCR{\Gamma_{\rm VCR}}

\begin{document}

\title{Relativity tests and their motivation}


\author{Ralf Lehnert}{
  address={Instituto de Ciencias Nucleares,
Universidad Nacional Aut\'onoma de M\'exico,
A.~Postal 70-543, 04510 M\'exico D.F., Mexico}
}

\begin{abstract}
Some motivations for Lorentz-symmetry tests
in the context of quantum-gravity phenomenology 
are reiterated. 
The description of the emergent low-energy effects 
with the Standard-Model Extension (SME) is reviewed. 
The possibility of constraining such effects
with dispersion-relation analyses of collider data 
is established.
\end{abstract}

\maketitle


\section{Introduction}

One of the principal cornerstones of present-day physics 
is the special theory of relativity. 
It was established at the beginning of the last century, 
and it has significantly transformed our understanding of space and time.
Despite substantial experimental scrutiny, 
there exists no credible observational evidence 
for deviations from relativity theory.
As a matter of fact, 
Lorentz invariance, 
the symmetry that underlies special (and general) relativity, 
has acquired a venerable status.
For example, 
it is an ingredient in most theoretical approaches 
to physics beyond the Standard Model and general relativity.

Nevertheless, 
the last decade has witnessed a revival of interest 
in experimental tests of Lorentz symmetry. 
This renewed interest stems primarily from the realization 
that a more complete theory 
unifying quantum physics and gravity 
is likely to affect the structure of spacetime at small distance scales. 
In fact, 
the majority of theoretical approaches to quantum gravity 
(although based on Lorentz symmetry) 
can accommodate minuscule deviations from special relativity 
in the ground state.
Such mechanisms for Lorentz violation exist,
for example, 
in string theory, 
spacetime-foam models, 
non-commutative field theory, 
and cosmologically varying scalars~\cite{lotsoftheory}. 

To identify and analyze present and near-future experimental tests of these ideas, 
a general framework for the description of Lorentz violation
at currently attainable energies is needed. 
Such a framework, 
known as the Standard-Model Extension (SME), 
has been developed in a series of papers~\cite{sme}. 
The SME is an effective field theory 
that incorporates practically all established physics 
in the form of the Standard-Model and general-relativity Lagrangians. 
In addition, 
it contains Lorentz- and CPT-breaking contributions
formed by contracting external, non-dynamical vectors and tensors 
with conventional particle and gravitational fields 
to form coordinate scalars.
The prescribed background vectors and tensors 
control the size and type of Lorentz and CPT violation 
and are amenable to experimental searches. 
The dominant Lorentz- and CPT-breaking effects 
are expected to arise 
from the power-counting renormalizable contributions to the SME.
This particular subset of the SME is called the minimal Standard-Model Extension (mSME);
in the past, 
it has provided the basis 
for numerous experimental~\cite{cpt07,kr} 
and theoretical~\cite{theory} investigations
of relativity theory.

One prediction of the SME 
that has been particularly popular 
is the modification of one-particle dispersion relations. 
The novel correction terms in such dispersion relations
typically involve the Lorentz-violating background 
contracted with certain powers of the particle's momentum, 
so that the Lorentz-breaking effects 
tend to be more significant at higher energies. 
Therefore, 
ultrahigh-energy cosmic rays (UHECRs) 
have traditionally been used 
for kinematical dispersion-relation tests of Lorentz symmetry~\cite{thres}.
However, 
the kinematics of an UHECR collision 
involves the (modified) dispersion relations of {\em all} involved particles 
including the primary, 
but the nature of the primary is often difficult to establish. 
For this reason, 
it is interesting to consider also Lorentz tests via particle collisions 
in a controlled laboratory environment 
at the cost of being confined to a lower-energy regime.
In what follows, 
we will therefore focus on dispersion-relation tests of special relativity
at highest-energy particle colliders~\cite{mesonexample}.

It turns out that Lorentz tests at particle accelerators
are particularly sensitive to the electron--photon sector of the mSME. 
Gravitational physics can be safely ignored. 
The mSME coefficients applicable in this context are
$(k_{F})^{\mu\nu\rho\lambda}$, $(k_{AF})^{\mu}$, 
$b^\mu$, $c^{\mu\nu}$, $d^{\mu\nu}$, and $H^{\mu\nu}$;
all of these non-dynamical background vectors and tensors 
are  spacetime constant.
It is important to note 
that the coefficients $c^{\mu\nu}$ and $\tilde{k}^{\mu\nu}\equiv(k_F)_\alpha{}^{\mu\alpha\nu}$ 
are physically equivalent 
in an electron--photon system. 
This equivalence arises 
because suitable coordinate rescalings 
freely transform the $\tilde{k}^{\mu\nu}$ and $c^{\mu\nu}$ parameters
into one another~\cite{km0102,collider}. 
From a physics perspective, 
this represents the fact 
that we may choose to measure distances 
with a ruler composed of electrons ($c^{\mu\nu}=0$), 
or with a ruler composed of photons ($\tilde{k}^{\mu\nu}=0$), 
or any other ruler ($c^{\mu\nu},\,\tilde{k}^{\mu\nu}\neq0$).
We exploit this freedom 
by selecting the specific scaling $c^{\mu\nu}=0$
(corresponding to an ``electron ruler'')
in intermediate calculations.
However, 
we state the final result in a scaling-independent 
(i.e., ``ruler-independent'') way
and reinstate the $c^{\mu\nu}$ coefficient 
for generality.

In principle, 
all of the above coefficients can contribute 
to the kinematics of the electron--photon vertex.
However, 
prior experimental bounds on Lorentz violation establish
that the dominant one of the above mSME coefficient is $k_F$~\cite{kr}. 
This coefficient causes a direction- and polarization-dependent speed of light~\cite{km0102}.
A number of its components 
have been tightly constrained with 
astrophysical polarimetry~\cite{km06}, 
Michelson--Morley tests~\cite{kr,newMM},
and Compton scattering~\cite{GRAAL}.
We will place limits on the $\ktr$ piece of $\tilde{k}^{\mu\nu}$,
which is its isotropic component~\cite{km0102}.
At the time of the analysis, 
it obeyed the weakest limits, 
so all other components of $\tilde{k}^{\mu\nu}$ components
can also be set to zero 
in this context.
An mSME calculation then shows 
that the photon's dispersion relation is modified:
in the presence of $\ktr$, 
it is given by \cite{km0102} 
\begin{equation}
E_{\gamma}^2-(1-\ktr)\vec{p}\!\phantom{.}^{2}=0\;.\label{eq:dispersion}
\end{equation}
Here, 
$p^{\mu}\equiv (E_{\gamma},\vec{p})$ is the photon's 4-momentum, 
and Eq.~(\ref{eq:dispersion}) holds at leading order in $\ktr$. 
This dispersion relation can be interpreted 
as a nontrivial isotropic refractive index $n$ of the vacuum:
\beq{analogy}
n=1+\ktr+{\cal O}\left(\ktr^2\right)\;.
\eeq
Note in particular 
that the physical speed of light is $(1-\ktr)$
(i.e., different from the usual $c=1$).
We also remark 
that the electron's dispersion relation 
$E(p)=\sqrt{m_e^2+p^2}$
remains unaltered
with our choice of coordinate scaling.
In what follows, 
we treat the two cases $\ktr<0$ and $\ktr>0$ separately
because they lead to different phenomenological effects. 

\section{Photon decay}

For negative $\ktr<0$,
photons travel faster than 
the maximal attainable speed (MAS) of electrons.  
This introduces photon instability: 
for photon energies $E_\gamma$ above the threshold 
\begin{equation}
E_{\rm pair}=\frac{2m_e}{\sqrt{\ktr(\ktr-2)}}=\sqrt{\frac{2}{-\ktr}}m_e
+\mathcal{O}\left(\sqrt{\ktr}\right)\; ,\label{gammathres}
\end{equation}
photon decay into an electron--positron pair 
is kinematically allowed~\cite{ks08,collider}.
This threshold condition 
can be established with the aid of the modified dispersion relation~(\ref{eq:dispersion}).
The leading-order decay rate of this process
is given by~\cite{ks08,collider}
\begin{equation}
\Gamma_{\rm pair}=\frac{2}{3}\,\alpha\, E_{\gamma}\,\frac{m_e^{2}}{E_{\rm pair}^{2}}\,
\sqrt{1-\frac{E_{\rm pair}^{2}}{E_{\gamma}^{2}}}\left(2+\frac{E_{\rm pair}^{2}}{E_{\gamma}^{2}}\right)\;,\label{pairrate}
\end{equation}
where $\alpha\simeq\frac{1}{137}$ denotes the fine-structure constant.
Note 
that this process is highly efficient.  
For example, 
a $40\,$GeV photon with energy 1\% above threshold 
would decay after traveling
about $30\, \mu$m.

The absence of such a photon-decay effect in nature 
can be used to obtain limits on negative values of $\ktr$ 
as follows. 
Suppose long-lived photons with a known energy $E_\gamma$ 
are observed to exist.
Such photons must essentially be below threshold $E_\gamma<E_{\rm pair}$,
for otherwise they would decay rapidly according to Eq.~(\ref{pairrate}).
Using the sub-threshold condition $E_\gamma<E_{\rm pair}$ in Eq.~(\ref{gammathres}) 
yields
\beq{subthresdecay}
E_\gamma\lsim\sqrt{\frac{2}{-\ktr}}m_e\quad\quad\textrm{or equivalently}
\quad\quad\ktr\gsim-2\,\frac{m_e^2}{E_\gamma^2}\;.
\eeq
It follows 
that stable photons with higher energies $E_\gamma$ 
give stronger bounds on negative values of $\ktr$.

Hadron colliders generate the highest-energy photons 
and therefore give tight Earth-based experimental limits on negative $\ktr$.  
This leads us to consider Fermilab's Tevatron $p\overline{p}$ collider 
with center-of-mass energies up to $1.96\,$TeV.
At the Tevatron,
isolated-photon production with an associated jet 
has been investigated with the D0 detector
because of its importance for QCD studies. 
In this context, 
the photon-energy bin 
at the ultraviolet end of the recorded spectrum
extended from $300\,$GeV to $400\,$GeV. 
We can therefore conservatively take $E_{\rm pair}>E_\gamma\simeq300\,$GeV.
With Eq.~(\ref{subthresdecay}), 
we then arrive at the constraint
\begin{equation} 
-5.8\times10^{-12}\lsim\ktr-\frac{4}{3}\,c^{00}\;, 
\label{decaybound}
\end{equation}
where we have reinstated the contribution of the electron's $c^{00}$ 
coefficient.

\section{Vacuum Cherenkov radiation}

For positive $\ktr>0$, 
photons are stable.
The speed of light is now $(1-\ktr)$.
Note in particular
that this speed is slower than the MAS of the electrons. 
In analogy to conventional electrodynamics inside a macroscopic medium,
this suggest a Cherenkov-type effect~\cite{cher}: 
charges moving faster than the modified speed of light $(1-\ktr)$ 
become unstable against the emission of photons.
Employing the Lorentz-violating dispersion relation~(\ref{eq:dispersion}), 
one can indeed establish 
that electrons with energies $E$ above the threshold
\begin{equation}
E_{\rm VCR}=\frac{1-\ktr}{\sqrt{(2-\ktr)\ktr}}\,m_e=\frac{m_e}{\sqrt{2\ktr}}+\mathcal{O}\left(\sqrt{\ktr}\right)\label{vcrenergy}
\end{equation}
emit Cherenkov radiation.
We remark 
that the threshold~(\ref{vcrenergy}) 
can alternatively be derived from 
the usual Cherenkov condition 
that the electron must be faster 
than the speed of light $(1-\ktr)$. 

Paralleling the photon-decay case in the previous section, 
we want to determine an experimental limit on $\ktr$ 
through the non-observation of vacuum Cherenkov radiation.  
To this end, 
we need to establish 
that this effect would be efficient enough
for a rapid deceleration of charges with $E>\EVCR$
to energies below threshold.
One can show 
that near $\EVCR$, 
the dominant deceleration process is
single-photon emission 
with an estimated rate of~\cite{Altschul:2008}
\begin{equation}
\GVCR=\alpha\, m_e^{2}\,\frac{(E-\EVCR)^{2}}{2E^{3}}\;,\label{vcrate}
\end{equation}
where $\alpha$ is again the fine-structure constant,
and $E$ denotes the electron energy, 
as before.  
A numerical evaluation of this expression indeed shows 
that the emission process is quite efficient, 
and above-threshold electrons would be extremely short-lived.

We can now employ the threshold condition~(\ref{vcrenergy}) 
together with the existence of high-energy electrons 
to place a limit on positive values of $\ktr$.
The observation of long-lived electrons at a known energy $E$ 
practically implies $\EVCR>E$. 
Using this information in Eq.~(\ref{vcrenergy}) gives
\beq{CherenkovLimit}
E\lsim\frac{m_e}{\sqrt{2\ktr}}\quad\quad\textrm{or equivalently}
\quad\quad\ktr\lsim\frac{1}{2}\,\frac{m_e^2}{E^2}\;.
\eeq
It is apparent 
that the limit on positive $\ktr$ gets tighter
with higher energies $E$ of the long-lived electrons.

The highest laboratory-frame electron energy 
at an Earth-based collider 
was reached at LEP, 
where the value $E_{\rm LEP}=104.5\,$GeV was attained.
Employing Eq.~\eqref{vcrate}, 
we can establish 
that if $\EVCR=104\;$GeV, 
electrons initially accelerated to $104.5\,$GeV 
would be rapidly decelerated 
by the emission of Cherenkov radiation 
to an energy below $\EVCR$ over a $1/e$ length of roughly $95\,$cm.  
The total energy loss due to the Cherenkov effect 
in such a scenario 
would far exceed the value allowed by measurements.  
With Eq.~\eqref{vcrenergy} at hand, 
the requirement 
that $\EVCR$ be greater than $104\;$GeV yields
\begin{equation}
\ktr-\frac{4}{3}\,c^{00} \leq 1.2\times 10^{-11}\;,\label{VCRbound}
\end{equation}
where we have again included the dependence on $c^{00}$ 
for generality.

\section{Conclusions}

We have shown 
that data from highest-energy particle colliders 
can be used to extract competitive limits
on isotropic Lorentz violation in the electron--photon system.
Combining the results~(\ref{decaybound}) and~(\ref{VCRbound}), 
we obtain the two-sided limit 
\begin{equation}
-5.8\times10^{-12}\leq\ktr-\frac{4}{3}\,c^{00}\leq1.2\times10^{-11}\;.
\label{bound}
\end{equation}
We remark 
that other aspects of collider physics
(namely modifications of synchrotron radiation) 
yield further improvements of this limit 
at the $10^{-15}$ level~\cite{Altschul:2009b}.


\begin{theacknowledgments}
The author wishes to thank the organizers
for the invitation to this stimulating meeting.
This work was funded in part 
by CONACyT under Grant No.\ 55310.  
\end{theacknowledgments}


\bibliographystyle{aipproc} 

\end{document}